\documentclass[12pt]{article}

\usepackage{amssymb,amsmath,amsfonts,eurosym,geometry,ulem,graphicx,caption,color,setspace,sectsty,comment,footmisc,caption,pdflscape,subfigure,array,hyperref,appendix,adjustbox}
\usepackage{url}
\usepackage{natbib}

\newcolumntype{L}[1]{>{\raggedright\let\newline\\arraybackslash\hspace{0pt}}m{#1}}
\newcolumntype{C}[1]{>{\centering\let\newline\\arraybackslash\hspace{0pt}}m{#1}}
\newcolumntype{R}[1]{>{\raggedleft\let\newline\\arraybackslash\hspace{0pt}}m{#1}}

\begin{document}

\begin{titlepage}
\title{
An open-source tool to assess the carbon footprint of research}
\author{Jérôme Mariette \thanks{Université de Toulouse, INRAE, UR MIAT, F-31320, Castanet-Tolosan, France} \and Odile Blanchard \thanks{Univiversité Grenoble Alpes, CNRS, INRAE, Grenoble INP, GAEL, 38000 Grenoble, France } \and Olivier Berné\thanks{Institut de Recherche en Astrophysique et Planetologie, Université de Toulouse, CNRS, CNES, UPS, Toulouse, France, 9 Av. du colonel Roche, 31028 Toulouse Cedex 04, France} \and Tamara Ben-Ari\thanks {UMR 211, INRAE, AgroParisTech, Université Paris-Saclay, F-78850 Thiverval-Grignon, France; CIRED, UMR 8568, F-94736 Nogent-sur-Marne, France} \and \ldots }
\date{\today}
\maketitle



\begin{center}
------------------------------------

\textbf{Version 1 of the paper : author list to be completed}

------------------------------------

\end{center}

\begin{abstract}

Research institutions are bound to contribute to greenhouse gas emission (GHG) reduction efforts for several reasons. First, part of the scientific community's research deals with climate change issues. Second, scientists contribute to students' education : they must be consistent and role models. Third the literature on the carbon footprint of researchers points to the high level of some individual footprints. In a quest for consistency and role models, scientists, teams of scientists or universities have started to quantify their carbon footprints and debate on reduction options. Indeed, measuring the carbon footprint of research activities requires tools designed to tackle its specific features. In this paper, we present an open-source web application, {\it GES 1point5}, developed by an interdisciplinary team of scientists from several research labs in France. {\it GES 1point5} is specifically designed to estimate the carbon footprint of research activities in France. It operates at the scale of research labs, {i.e., \it  laboratoires}, which are the  social structures around which research is organized in France and the smallest decision making entities in the French research system. The application allows French research labs to compute their own carbon footprint along a standardized, open protocol. The data collected in a rapidly growing network of labs will be used as part of the {\it Labos 1point5} project to estimate France's research carbon footprint. We expect that an international adoption of {\it GES 1point5} (adapted to fit domestic specifics) could contribute to establishing a global understanding of the drivers of the research carbon footprint worldwide and the levers to decrease it.

\textbf{Availability and implementation:} GES 1point5 is available online for French research labs at \url{http://labos1point5.org/ges-1point5} and source code can be downloaded from the GitLab platform at \url{https://framagit.org/labos1point5/l1p5-vuejs}.

\vspace{0in}
\vspace{0in}

\bigskip
\end{abstract}
\setcounter{page}{0}
\thispagestyle{empty}
\end{titlepage}
\pagebreak \newpage


\doublespacing

\section{Introduction}

When signing the Paris Agreement, France committed to drastically reducing its domestic greenhouse gas (GHG) emissions within the next decades. All sectors are expected to contribute to this mitigation effort, including the public sector. Although it is expected that the research sector's direct contribution to domestic GHG emissions is small, research institutions are bound to contribute to these reduction efforts for several reasons. First, part of the scientific community's research deals with climate change issues; their results point to the need to mitigate and adapt to climate change. The whole scientific community cannot ignore those results. Second, scientists contribute to students' education and often stand at the forefront of the discussions on climate change and its impacts on ecosystems and societies : they must be consistent and be role models. Third the literature on the research sector's GHG emissions points a high footprints of some scientists \citep{na20,spi13,fox09,gre08}.

At the international level, only a small number of studies report on estimates of the carbon footprint of the research sector. These studies either focus on one single entity such as a research lab or a university \citep{gue13, wyn19}, on a specific event such as a conference \citep{spi13, des16, str15, klo20}, on a single source of GHG emissions such as air travel \citep{cie19}, or on a specific research project \citep{ach13,bar20}. When more comprehensive, these estimates relate only to one single year \citep{gue13}. Furthermore, all of these studies rely on the development of ad-hoc methodologies relying on a specific set of emission factors (\textit{i.e.}, the parameters used to convert activity levels into GHG emissions). This point prevents from any comparison between carbon footprints and highlights the need for a dedicated and standardized application.

Several tools are actually freely available. Some are designed to estimate individual or households' GHG emissions and may be bounded to lucrative carbon offset schemes\footnote{\url{https://co2.myclimate.org/en/}}\footnote{\url{https://www.carbonfootprint.com/}}\footnote{\url{https://offset.climateneutralnow.org/footprintcalc}}\footnote{\url{https://nosgestesclimat.fr/simulateur/bilan}}\footnote{\url{https://www.goodplanet.org/fr/calculateurs-carbone/particulier/}}. The GHG Protocol\footnote{\url{https://ghgprotocol.org/}} offers a set of worksheets dedicated to industries, businesses, countries or cities. The University of New Hampshire has developed SIMAP$^{\mbox{\scriptsize{\textregistered}}}$\footnote{\url{https://www.unh.edu/sustainability/research/campus-calculator-tools}}, a standardized software to address the specific needs of North American education institutions, colleges, and universities. In France, \textit{Bilan Carbone}$^{\mbox{\scriptsize{\textregistered}}}$\footnote{\url{https://www.associationbilancarbone.fr/les-solutions/}} is a generic tool developed as a set of worksheets and requires its users to pay for a licence and a training program. None of these tools are specifically designed to estimate the carbon footprints of research activities.

In this paper, we present a free and open source standardized web application, \textit{GES 1point5}. It has been developed by an interdisciplinary team of engineers and researchers who work in various research labs in France and interact in the non governmental organization \textit{Labos 1point5}\footnote{\url{https://labos1point5.org/}}. The tool allows any research lab to (i) estimate the emissions relating to the energy consumption and refrigerant gases of its buildings, and to its members' travels (ii) easily highlight, via a graphic interface, the main GHG emissions drivers of a research lab, and (iii) design emission reduction actions and evaluate their impact over time. Because they rely on a standardized protocol, completed \textit{GES 1point5}-based GHG inventories can  be compared. This enables to make progress in our understanding of the drivers of the research footprint (e.g., disciplinary or geographic). At the national scale, the data collected will allow to estimate the carbon footprint of the French public research and thus support the exploration of evidence-based emission reduction strategies.

The paper is organised as follows: we first describe a few specifics of the French research system and the goals of \textit{GES 1point5}; we then describe \textit{GES 1point5} methodology and its implementation; the fifth section provides an illustration of \textit{GES1 point5} outputs, namely  the GHG inventory and carbon footprint of a fictitious research lab; future developments and research perspectives are addressed in the sixth section, before the conclusion.

\section{Context and goals}

The French research system encompasses several types of institutions \footnote{\url{https://www.sciencemag.org/careers/2006/04/finding-your-way-around-french-research-system}}: national research institutes (such as CNRS, INRAe, CEA, IRD, \ldots), semi-private research institutions (such as CIRAD, Ifremer, etc.) and universities. These institutions take part in social structures called \textit{laboratoires} (referred to as ``research labs'' in the following sections). Their financial contributions to the operation of research labs may be multiple, \textit{e.g.} they may pay the salaries or stipends of its members, provide fixed assets such as buildings or infrastructures, pay for resources such as supplies, electricity, etc. A typical \textit{laboratoire} benefits from the involvement of several institutions or universities. These research labs comprise between tens and at most a few hundred members, and occupy one or several buildings. A \textit{laboratoire} usually focuses on a specific scientific field. France counts over 1000 public research labs overall.

In this context, the implementation of \textit{GES 1point5} was initiated to provide the French scientific community with a free and open source tool able to estimate the carbon footprint of a research lab using a standardized methodology. {\it GES 1point5} operates at the scale of research labs for several reasons. {\it GES 1point5} is part of {\it Labos 1point5}\footnote{\url{labos1point5.org}}, a large interdisciplinary project in France that aims at fueling a bottom-up momentum by mobilizing members of the research community as opposed to decision-makers in the academic bureaucracy. Research labs are the smallest human scale entities in the research system. They rely on independent and collective decision-making processes: experimental designs, as well as access to research facilities are decided and managed at the research lab scale. Consequently, a number of decisions relevant to the reduction of the carbon footprint can be made at the scale of research labs.

Sharing a common methodology allows to compare and aggregate results to analyse and answer research questions at the research lab scale but also at smaller and larger scales. At the research lab scale, the carbon footprint can be analysed to understand (i), the main emission sources and the relative contribution of each activity, (ii), the relative contribution of its members according to their professional status, and (iii), the emission dynamics over time relative to the mitigation actions that may have been implemented. Indeed, \textit{GES 1point5} is meant to be a decision support tool for mitigation actions and experiments. For example, to reduce emissions from professional travels, different mitigation decisions or policies may be tested, such as an internal carbon tax or individual emission quotas.

At a smaller scale, \textit{GES 1point5} may also allow to estimate the carbon footprint of specific research projects or specific teams within the lab, that of PhD theses or even of conferences.

At a larger scale, another objective with \textit{GES 1point5} is to create a dataset which is a collection of the GHG footprints of a large number of research labs in France in a large array of disciplines. Combining this dataset with statistical analyses, it will be possible to perform an extrapolation to assess the overall carbon footprint of French public research, and to describe the distribution of sources of emissions across regions and disciplines. 

Coupling the analyses performed at different scales with the experiences led by research labs will contribute to the construction of robust emission reduction scenarios at the global level and recommendations in terms of public research policy.

\section{\textit{GES 1point5} methodology}

\textit{GES 1point5} complies with and refers to the French legislation \citep{mee16}. When estimating the carbon footprint of an entity, it is necessary to precisely define its scope \textit{i.e.}  the GHGs considered and the emission sources included, as well as to mention the emission factors chosen. The next two subsections present \textit{GES 1point5} features in these respects.

\subsection{Scope}

As required by the French legislation, the GHGs considered in \textit{GES 1point5} are those of the Kyoto Protocol. In terms of emission sources,\textit{GES 1point5} takes into account those that are common and most often critical in research labs: buildings through energy consumption, air conditioning and refrigeration processes, daily commutes and professional travels due to the use of cars, boats or planes to attend meetings or harvest field data.

\subsection{Emission factors}
\label{sec:emissions-factor}

An emission factor represents the amount of GHG emissions generated by a unit of activity. The emission factors included in \textit{GES 1point5} mainly stem from the ADEME database\footnote{\url{https://www.bilans-ges.ademe.fr/en/accueil/}} that is the backbone of the French legislation. When an emission factor is missing from the ADEME database, the application refers to the most recent available studies. This is the case for example for an electric bike \citep{ave15} or for an electric scooter \citep{arc19}. Furthermore, the \textit{GES 1point5} team has created customized emission factors to take into account specific research lab activities. For example, the application includes an emission factor for research campaigns at sea, based on ad-hoc estimates carried out by some laboratories.

\textit{GES 1point5} also corrects some of the ADEME database inconsistencies. For example, in its current version\footnote{version 19.0}, the ADEME database does not include manufacturing emissions in the emission factor per kilometer of a gasoline car, whereas it does for hybrid and electric cars. As manufacturing emissions are significant when considering the total emissions of a car, these emissions are included in \textit{GES 1point5} for all types of cars, using factors that were provided in a previous version of the ADEME database.
More generally, \textit{GES 1point5} includes manufacturing emission factors for all vehicles when these emissions are significant, \textit{i.e.}, cars, buses, coaches, trains, streetcars and subways. The application includes manufacturing emissions drawn from previous versions of the ADEME database and adjusted proportionately to the weight they represented in the total emission factor of the previous versions.

All methodological choices are thoroughly detailed in the methodology section of the tool available online.

\section{Implementation}
\label{sec:implementation}

\textit{GES 1point5} has been implemented as a set of components using VueJS\footnote{\url{https://vuejs.org/}} and django\footnote{\url{https://www.djangoproject.com/}} frameworks. The application uses input information that can be gathered reasonably easily (provided support is granted by the administration) and converts it into a GHG footprint. For a given number of emission sources, \textit{GES 1point5} converts GHG-emitting activity levels into CO$_2$${e}$ (carbon dioxyde equivalent) using emission factors as described in Section~\ref{sec:emissions-factor}. 

\begin{figure}[!tpb]
  \centering
  \begin{tabular}{cc}
    \includegraphics[width=1\linewidth]{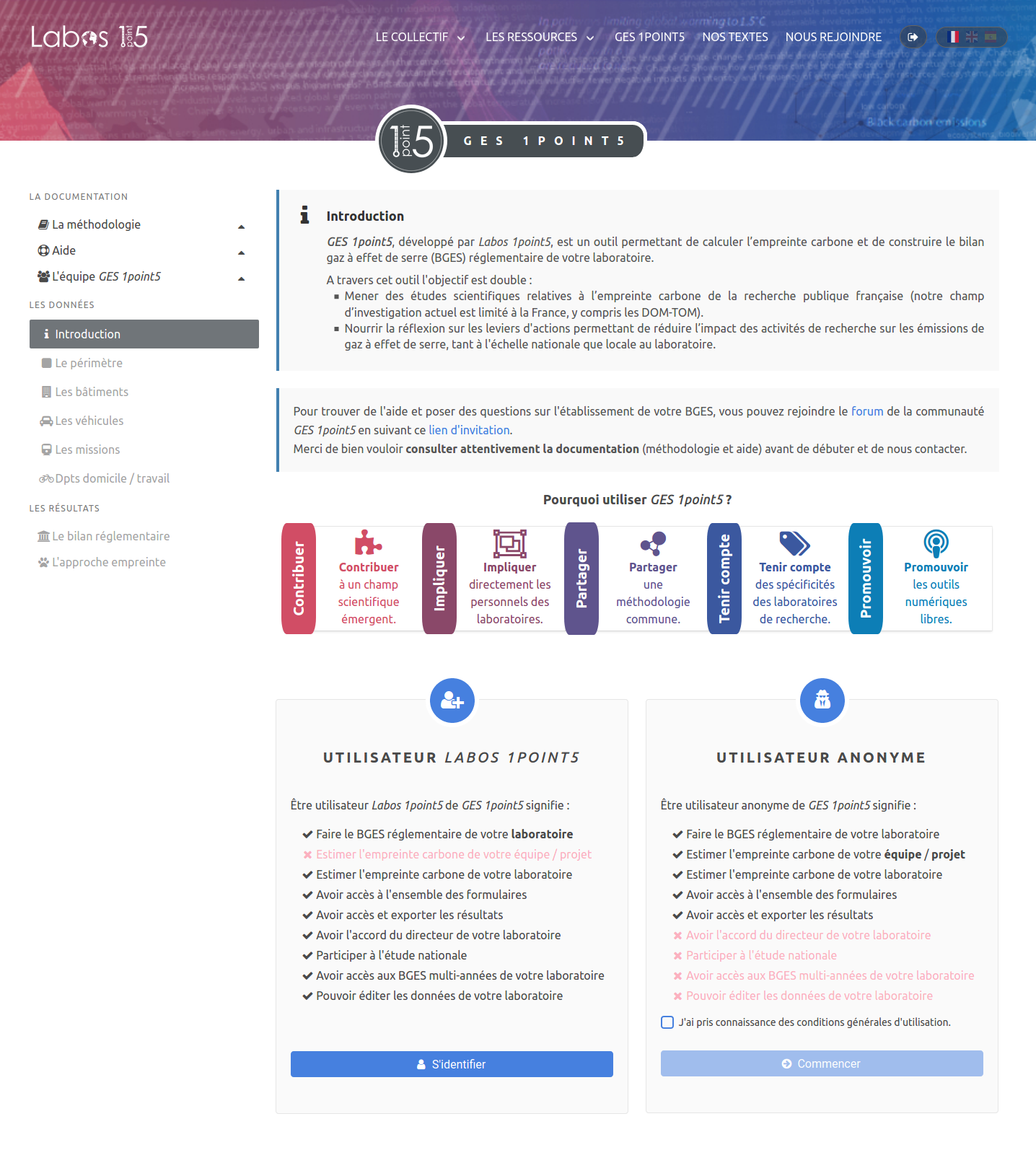}
  \end{tabular}
  \caption{The \textit{GES 1point5} welcome page offers its user the opportunity to estimate GHG emissions anonymously or using an authenticated account. The application menu, on the left, allows to navigate between documentation pages and the different emission sources forms.}
  \label{fig::welcome-page}
\end{figure}

From its welcome page (Figure~\ref{fig::welcome-page}), the application offers its users the opportunity to estimate GHG emissions anonymously or using an authenticated account. In the latter case,  \textit{GES 1point5} allows to store input and output data relating to the research lab emissions. To gather the data required, the application provides a set of forms and routines briefly described below.

\subsection{Inputs}

\begin{figure}[!tpb]
  \centering
  \begin{tabular}{cc}
    \includegraphics[width=1\linewidth]{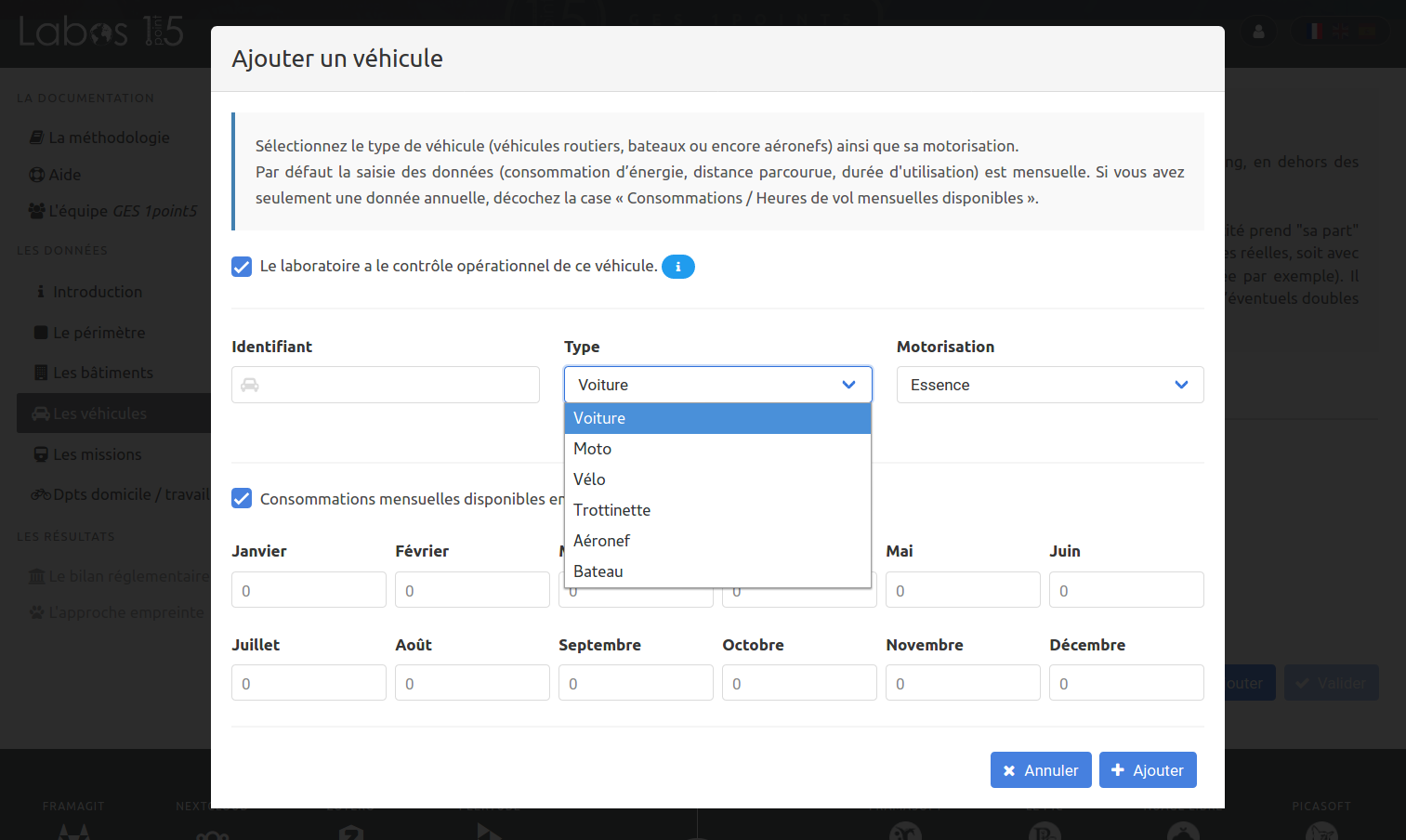}
  \end{tabular}
  \caption{Form to add a new vehicle to the research lab's inventory. The vehicle can be either a car, a motorbike, a bike, a scooter, an aircraft or a boat. The form requires to define the vehicle motorisation and its annual consumption.}
  \label{fig::vehicules-form}
\end{figure}

\begin{itemize}
    \item General information : year of the GHG inventory ; number of lab members according to their position (e.g. research members, technicians, administrative staff, PhD students,post-doctoral fellows). 
    \item Data on buildings : floor area ; consumption of electricity, heat, and refrigerant gases ; specifics related to the generation of electricity, when applicable (e.g. use of solar panels). 
    \item Data on transportation modes operated by the lab : type (e.g. car, truck, aircraft) ; type of fuel ; power, distances travelled, number of hours of operation when applicable. Figure~\ref{fig::vehicules-form} presents the form dedicated to add a new vehicle and entering its energy consumption.
    \item Commutes : a standardized survey dedicated to collect lab members' daily commutes is provided by \textit{GES 1point5}, using the \textit{Framaform} web application provided by \textit{framasoft}, a not-for-profitorganization that makes the safe handling of personal data overarching. The survey is to be sent to all lab members. Once the survey period is over, results can be exported from \textit{Framaform} and directly imported into \textit{GES 1point5}.
    \item Professional travel : the raw data are extracted from the information systems of the various research institutions that pay for the travels of the lab members. For each trip, they comprise information on the date, the departure and destination places (cities, countries), the travel modes, the travel purpose, the status of the lab member. These data are imported as a .tsv file into \textit{GES 1point5}, Table~\ref{tab::professional-travels-format} defines the required format.
\end{itemize}

\begin{table*}[h!]
\caption{Column description of the tsv file accepted by \textit{GES 1point5} to import professional travels.\label{tab::professional-travels-format}}

{\footnotesize
\begin{adjustbox}{width=1\textwidth}
\begin{tabular}{|p{5cm}|p{8cm}|}
\hline

Column ID & Description \\ \hline
Trip number & One line per leg of the trip ; the trip number is a simple sequence number (from 1 to n) which allows to gather all the legs of the same trip. If the return trip is different from the outward trip, two lines must be used. \\ \hline
Departure date & A date in dd/mm/yyyy format. \\ \hline
Departure city & The departure city name. This field is used to obtain the departure city GPS coordinates using the \textit{geonames}\footnote{\url{https://www.geonames.org/}} database. \\ \hline
Departure country & The departure country or its ISO3166 code. \\ \hline
Destination city &  The destination city name. This field is used to obtain destination city GPS coordinates using the \textit{geonames} database. \\ \hline
Destination country &  The destination country or its ISO3166 code. \\ \hline
Travel mode & This field can take a value among [ "Avion", "Train", "Voiture", "Taxi", "Bus", "Tramway", "RER", "Métro", "Ferry" ]. \\ \hline
Number of people in the car & In case of a trip in a car or a taxi, number of people in the vehicle \\ \hline
One way / return & If the outward and return trips are identical, enter "OUI", otherwise "NON". \\ \hline 
Travel purpose (optional) & This optional field allows the application to perform emissions statistics based on the travel purpose. This field can take a value among [ "Etude terrain", "Colloque-Congrès", "Séminaire", "Enseignement", "Collaboration", "Visite", "Administration de la recherche", "Autre" ]. \\ \hline
Agent status (optional) & This optional field allows the application to perform emissions statistics based on agents' statuses. This field can take a value among [ "Chercheur.e-EC", "ITA", "Doc-Post doc", "Personne invitée" ]. \\ \hline
\end{tabular}
\end{adjustbox}}
\end{table*}

\subsection{Outputs}

As previously mentioned, \textit{GES 1point5} complies with the French legislation, which itself abides by the GHG Protocol standard \citep{wri04}. The application thus provides the regulatory table covering scope 1 (direct emissions from owned or controlled sources), scope 2 (indirect emissions from the generation of purchased electricity, heating and cooling), as well as some emission categories among all other indirect emissions of scope 3. The resulting table, presented in Figure~\ref{fig::BGES-reglementaire} can be downloaded in csv format.

On top of displaying GHG emissions distribution within the three regulatory scopes, \textit{GES 1point5} provides a user-friendly synthetic representation of the research lab's carbon footprint. It helps the user to analyse which emission sources most impact the overall carbon footprint of the research lab, and to decide which actions to implement in order to mitigate the lab's GHG emissions. For example, in this representation, the buildings' direct and indirect emissions generated from the heating and cooling systems are aggregated, whereas they are split among the three scopes in the regulatory display. Similarly, the travel carbon footprint aggregates GHG emissions of vehicles, members' daily commutes and professional travels. In the regulatory table, these emissions are reported in scopes 1 and 3. Some examples of this operational representation of the carbon footprint are reported in Figure~\ref{fig::carbon-footprint} and \ref{fig::carbon-footprint-travel}. All the figures provided can be downloaded in multiple formats, such as png, jpeg, svg or pdf.

\section{Illustration}

This section provides and discusses results of the 2019 greenhouse gas inventory of a fictitious research lab named Cogitamus. The lab comprises 80 members distributed as follows : 14 researchers, 24 associate professors, 17 engineers or administrative staff and 25 PhD students or postdoctoral fellows. Cogitamus occupies one building shared with another lab and 60\% of the total floor space, 3~300 $m^2$. In 2019, the building consumed 200~000 kWh PCI from the Toulouse Canceropole urban heating network, 120~000 kWh of electricity and 0.3 kg of the R32 cooling gas. The lab owns one diesel car which traveled 12~000 km in 2019. For a full reproducibility of the results presented in this section, the commuting survey results\footnote{\url{https://cloud.le-pic.org/s/HaZWEae7ef22KMy}} and the professional travel file\footnote{\url{https://cloud.le-pic.org/s/msGYms2w7kqCNqJ}}  used are freely available.

\begin{figure}[!tpb]
  \centering
  \begin{tabular}{cc}
    \includegraphics[width=1\linewidth]{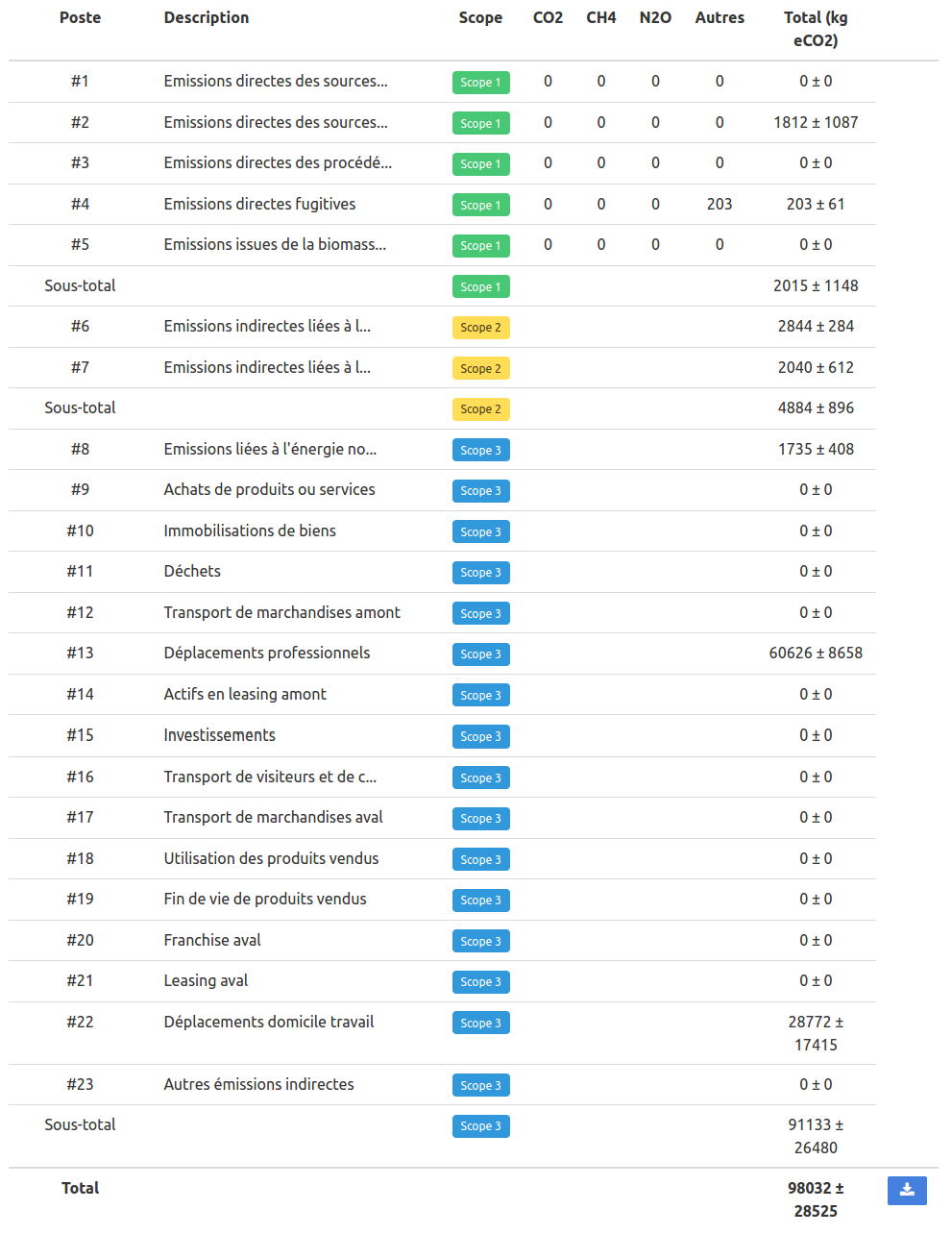}
  \end{tabular}
  \caption{Illustration of a regulatory table obtained with the \textit{GES 1point5} tool for a fictitious research lab, presenting the emissions distributed among the three scopes of the GHG Protocol standard. The translation to English is given in Table~\ref{table::traduction} }
  \label{fig::BGES-reglementaire}
\end{figure}

\begin{figure}[!tpb]
  \centering
    \includegraphics[width=1\linewidth]{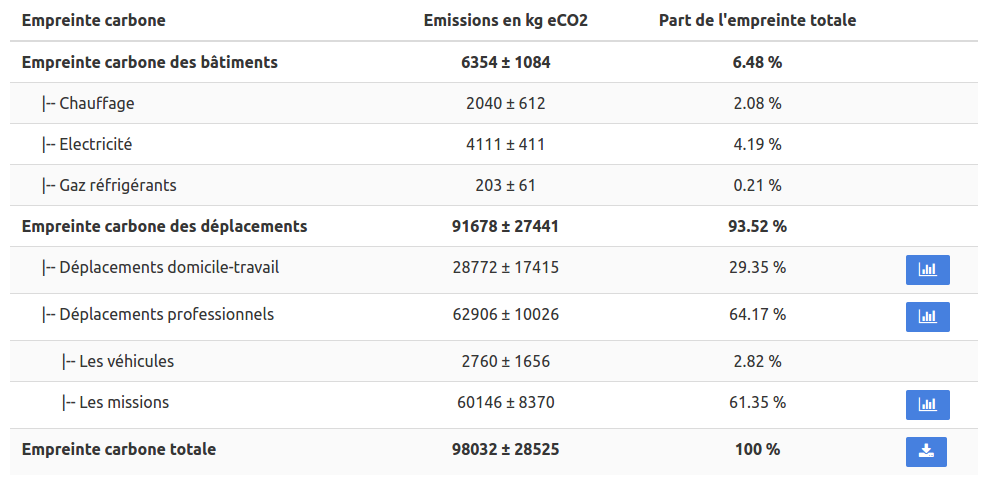}
    \includegraphics[width=1\linewidth]{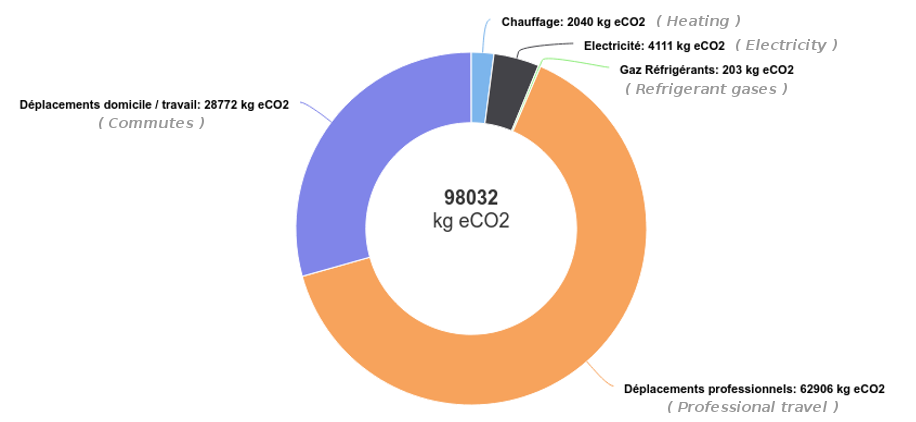}
  \caption{Illustration of the carbon footprint information provided as an output of \textit{GES 1point5} : table (upper panel) and of the pie chart (lower panel), showing the distribution of emissions in kg CO$_2$e. Translation of the terms is provided in Table~\ref{tab::traduction2}.}
  \label{fig::carbon-footprint}
\end{figure}

\begin{figure}[!tpb]
  \centering
  \begin{tabular}{cc}
    \includegraphics[width=1\linewidth]{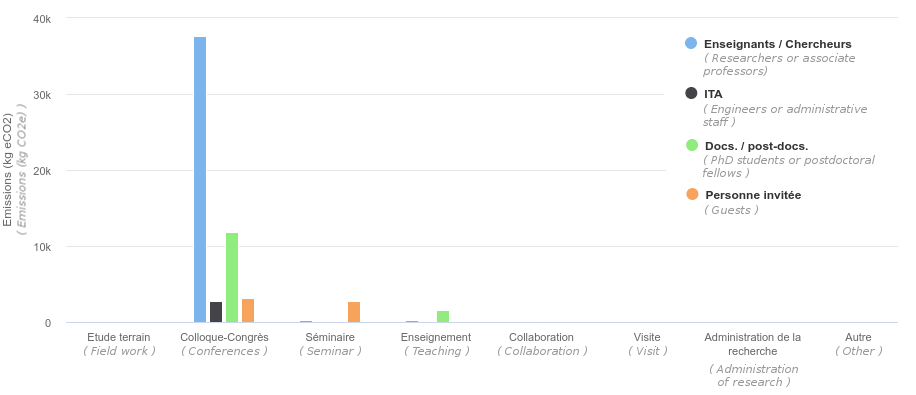}
  \end{tabular}
  \caption{Illustrative distribution of the professional travel carbon footprint provided by \textit{GES 1point5} according to travel purposes and agent status.}
  \label{fig::carbon-footprint-travel}
\end{figure}

Figures~\ref{fig::BGES-reglementaire}, \ref{fig::carbon-footprint} and \ref{fig::carbon-footprint-travel} illustrate the GHG emissions inventory that complies with the GHG Protocol standard, the user-friendly carbon footprint representation, graphs excerpted from \textit{GES 1point5}, respectively. More specifically, figure~\ref{fig::BGES-reglementaire} and \ref{fig::carbon-footprint} display the total emissions of the Cogitamus lab (98~032 $\pm$ 28~525 kg CO2e) from two different perspectives, \textit{i.e.},the regulatory GHG inventory table and the detailed carbon footprint. Figure~\ref{fig::carbon-footprint} (top) shows that these emissions are mainly driven by professional travels (64\%) followed by members' daily commutes (29\%) and electricity (4\%). Moreover, Figure~\ref{fig::carbon-footprint} (bottom) illustrates that emissions generated by professional travels are dominated by congress attendees' travels and more specifically by researchers' travels to these congresses.

Even if these results are fictitious, they highlight the benefits of using \textit{GES 1point5} to point to the need to focus primarily on travel emissions for an efficient mitigation action, even if reducing the buildings' emissions may be implemented in case of low-hanging fruits. Such results become even more valuable when analysing the changes in emissions over the years and evaluating the efficiency of the mitigation actions implemented.

\section{Discussion and perspectives}
 
\textit{GES 1point5} is a first step in a larger endeavour to facilitate and inform emission reductions from research activities in France. Estimating the carbon footprints is a crucial step to mitigate research labs' GHG emissions as it brings clearer information compared to that stemming from scopes 1, 2 and 3 of the GHG Protocol standard. The carbon footprints point to the relative weights of the main building blocks of the research labs’ GHG emissions. For example, if the buildings' carbon footprint of a research lab is very low compared to its travel carbon footprint, mitigation actions should mainly focus on the travel emissions - even if reducing the buildings' emissions may be implemented in case of low-hanging fruits. Similarly, within the travel carbon footprint, if the emissions from the professional travels overshadow the emissions from the staff and faculty commutes, mitigation actions should primarily focus on professional travels - even if raising awareness on the emissions from the commutes and acting to reduce them should also be addressed.
\textit{GES 1point5} helps to design and monitor mitigation actions in other respects, too. At the research lab level, while storing the GHG inventories year after year, it paves the way for analyses of the lab’s emissions dynamics, define and implement action plans, monitor progress and define new action plans in the long run. At the national level, statistical analyses can be carried out based on the aggregated data-set of all labs, and help to design public policies towards the mitigation of GHG emissions of the public research sector.

This first version of \textit{GES 1point5} presented in this paper focuses on the common emission sources that are often critical in research labs. However, other potentially important emission sources need to be soon assessed, among which internal information technology (IT) emissions, and emissions linked to purchased goods.

The estimation of the carbon footprint of IT systems internal to the research labs will be developed in \textit{GES 1point5} in future work. It will rely on the EcoDiag\footnote{\url{https://ecoinfo.cnrs.fr/ecodiag/}} database that provides manufacturing and transportation emission factors for multiple types of IT devices. In this case, electricity consumption emissions will not be estimated as they are already included in the emissions of the labs' buildings .

The estimation of the carbon footprint of purchased goods is complex as emission factors do not exist for all the categories of products that research labs can purchase. For example, it is common to use specific chemical solvents in biology labs, with only very little information on the manufacturing and supply chains. Various methodologies are currently under study and \textit{Labos 1point5} will thrive to consider the right trade-off between comprehensiveness and representativeness. 

\textit{GES 1point5} will later on be complemented by the estimation of the GHG emissions from the use of observational and experimental infrastructures (e.g. particle accelerator, telescope).

\section{Conclusion}

The research sector must reduce the carbon footprint of its activities along with all other sectors, in order for France to reach the goal of the Paris Agreement and, more indirectly, to sustain the bond between science and society. A first step naturally pertains to estimating current emissions level. In this context, a standardized tool is essential as it paves the way for comparisons of carbon footprints, statistical analyses, fruitful dialogues, coordinated mitigation strategies and reporting of mitigation actions results in a consistent way.

As an open-source software, \textit{GES 1point5} is freely accessible to any research lab in the world. Foreign research labs may have to adjust emission factors to their country when using the application, thus enabling comparisons between research labs worldwide. \textit{GES 1point5} opens up a very broad range of international research perspectives and initiates a strategy to discuss on the best ways to reduce research emissions worldwide.


\singlespacing
\setlength\bibsep{0pt}
\bibliographystyle{unsrtnat}
\bibliography{biblio}

\appendix

\section{Appendix tables}

\begin{table*}
\caption{Translation of the sources of emissions used in \textit{GES 1point5} and presented in Figure~\ref{fig::BGES-reglementaire}. The three scopes correspond to those of the GHG protocol. \label{table::traduction}}
\begin{adjustbox}{width=1\textwidth}

\begin{tabular}{|p{0.5cm}|p{7cm}|p{7cm}|}
\multicolumn{3}{c}{{\bf Scope 1}}\\ \hline
1 & Émissions directes des sources fixes de combustion & Direct emissions from stationary combustion sources \\ \hline
2 & Émissions directes des sources mobiles à moteur thermique & Direct emissions from mobile combustion sources \\ \hline
3 & Emissions directes des procédés hors énergie & Direct emissions from non-energy processes  \\ \hline
4 & Émissions directes fugitives & Direct fugitive emissions \\ \hline
5 & Emissons issues de la biomasse (sols et forêts) & Emissions from biomass (soils, forests) \\ \hline
\multicolumn{3}{c}{{\bf Scope 2}}\\ \hline
6 & Émissions indirectes liées à la consommation d’électricité & Indirect  emissions from purchased electricity  \\ \hline
7 & Emissions indirectes liées à la consommation de vapeur, chaleur ou froid & Indirect emissions from steam, heating or cooling \\ \hline
\multicolumn{3}{c}{{\bf Scope 3}}\\ \hline
8 & Emissions liées à l’énergie non incluse dans les catégories ``émissions directes de GES'' et ``émissions indirectes de GES associées à l’énergie'' & Emissions linked to energy non included in the ``Direct emissions'' and ``Indirect emissions associated with energy'' categories   \\ \hline
9 & Achats de produits ou services & Purchased goods and services \\ \hline
10 & Immobilisation des biens & Fixed assets \\ \hline
11 & Déchets & Waste \\ \hline
12 & Transport de marchandises amont & Transportation of goods upstream \\ \hline
13 & Déplacements professionnels & Employee business travel \\ \hline
14 & Actifs en leasing amont & Leased assets upstream \\ \hline
15 & Investissements & Investments \\ \hline
16 & Transports de visiteurs et de clients & Customer and visitor travel \\ \hline
17 & Transport de marchandise aval & Transportation of goods downstream \\ \hline
18 & Utilisation de produits vendus & Use of sold products \\ \hline
19 & Fin de vie des produits vendus & End of life of sold products \\ \hline
20 & Franchise aval & Franchises downstream \\ \hline
21 & Leasing aval & Leased assets downstream \\ \hline
22 & Déplacements domicile travail & Employee commuting \\ \hline
23 & Autres émissions indirectes & Other indirect emissions \\ \hline
\end{tabular}
\end{adjustbox}
\end{table*}

\begin{table*}
\caption{Translation of the terms used in the carbon footprint table presented in Fig.~\ref{fig::carbon-footprint} \label{tab::traduction2}}

\newpage

\begin{adjustbox}{width=1\textwidth}

\begin{tabular}{|p{5cm}|p{5cm}|p{5cm}|}
\hline
Carbon footprint & Emissions in kg CO$_2$e & Share of total fooprint \\ \hline
Carbon footprint of the buildings & &\\ \hline
- Heating & &\\ \hline
- Electricity & &\\ \hline
- Refrigerant gases & &\\ \hline
Travel carbon footprint & &\\ \hline
- Commutes & &\\ \hline
- Vehicles & &\\ \hline
- Professional travel & &\\ \hline

\end{tabular}
\end{adjustbox}
\end{table*}

\onehalfspacing

\end{document}